\documentclass[11pt,preprint,showpacs,preprintnumbers,amsmath,amssymb]{revtex4}

\usepackage{setspace}
\usepackage{graphicx}
\usepackage{dcolumn}
\usepackage{bm}
\usepackage{multirow}
\usepackage{graphicx}
\usepackage{amssymb}
\usepackage{float}
\usepackage{CJK}
\usepackage{caption}
\usepackage{rotating}
\usepackage{subfigure}
\usepackage{relsize}
\usepackage{diagbox}
\usepackage{epstopdf}
\usepackage{slashed}

\usepackage{exscale}
\usepackage{relsize}
\usepackage{dcolumn}
\usepackage{bm}
\usepackage{diagbox}
\usepackage[T1]{fontenc}
\usepackage{caption}

\begin{document}

\title{The masses and decay widths of the $S$-wave $\Lambda_c\bar{\Lambda}_c$ bound states}

\author{Shi-Ji Cao }
\affiliation{\scriptsize{Physics Department, Ningbo University, Zhejiang 315211, China}}

\author{Jing-Juan Qi }
\affiliation{\scriptsize{College of Information and Intelligence Engineering, Zhejiang Wanli University, Zhejiang 315101, China}}

\author{Zhen-Yang Wang \footnote{Corresponding author, e-mail: wangzhenyang@nbu.edu.cn}}
\affiliation{\scriptsize{Physics Department, Ningbo University, Zhejiang 315211, China}}

\author{Xin-Heng Guo \footnote{Corresponding author, e-mail: xhguo@bnu.edu.cn}}
\affiliation{\scriptsize{College of Nuclear Science and Technology, Beijing Normal University, Beijing 100875, China}}

\date{\today}

\begin{abstract}
In this work, we investigate possible bound states of the $\Lambda_c\bar{\Lambda}_c$ system in the Bethe-Salpeter formalism in the ladder and instantaneous approximations. By numerically solving the Bethe-Salpeter equation, we confirm the existence of $\Lambda_c\bar{\Lambda}_c$ bound states with quantum numbers $J^{PC}=0^{-+}$ and $J^{PC}=1^{--}$. We further investigate the partial decay widths of the $\Lambda_c\bar{\Lambda}_c$ bound states into $N\bar{N}$, $D\bar{D}$, $D\bar{D}^\ast$, $D^\ast\bar{D}^\ast$, $\pi\bar{\pi}$, and $K\bar{K}$. Our results indicate that the decay width of the $\Lambda_c\bar{\Lambda}_c$ bound state with $J^{PC}=1^{--}$ is much larger than that with $J^{PC}=0^{-+}$, and among their decay channels, the $D\bar{D}^\ast$ final state is the main decay mode. We suggest experiments to search for the $\Lambda_c\bar{\Lambda}_c$ bound states in the $D\bar{D}^\ast$ final state.
\end{abstract}

\pacs{******}

\maketitle
\section{Introduction}
\label{intro}
Exotic hadrons have become a major focus in both experimental and theoretical research due to their potential to reveal fundamental properties of strong interactions. With the efforts of experimental and theoretical studies, numerous heavy flavour exotic hadrons have been discovered, such as $X(3872)$ \cite{Belle:2003nnu}, $Z_c(3900)$ \cite{BESIII:2013ris}, and $P_c$ \cite{LHCb:2015yax,LHCb:2019kea}. These exotic hadrons are believed to have four or five quarks, their masses are typically located near the thresholds of either two mesons or one meson and one baryon. Recently, the LHCb, CMS, and ATLAS collaborations observed several exotic structures in the di-$J/\psi$ invariant mass spectra \cite{LHCb:2020bwg,ATLAS:2023bft,CMS:2023owd}, including the $X(6200)$, $X(6600)$, $X(6900)$, and $X(7200)$, which are candidates for fully-charmed tetraquark states. In addition to tetraquark and pentaquark states, it is natural to extend the research to the study of heavy flavour hexaquark states.

\begin{figure}[ht]
\centering
    \rotatebox{0}{\includegraphics*[width=0.4\textwidth]{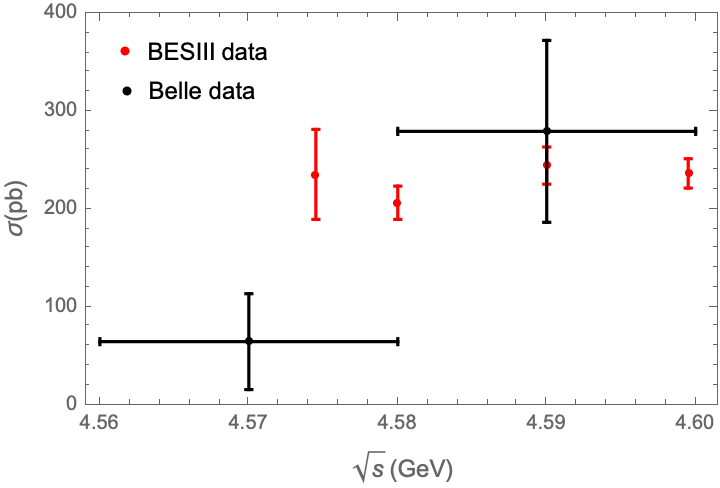}}
    \caption{Cross section of $e^+e^-\rightarrow\Lambda_c\bar{\Lambda}_c$
obtained by BESIII and Belle.}
  \label{data}
\end{figure}

In 2017, BES$\text{\uppercase\expandafter{\romannumeral3}}$ Collaboration carried out precision measurements of the cross section for the $e^+e^-\rightarrow\Lambda_c\bar{\Lambda}_c$ process at four energy points just above the $\Lambda_c\bar{\Lambda}_c$ threshold \cite{BESIII:2017kqg}. These measurements, depicted in Fig. \ref{data}, reveal an intriguing pattern: a discernible non-zero cross section close to the $\Lambda_c\bar{\Lambda}_c$ threshold and an almost constant profile. This contrasts sharply with the earlier results obtained by the Belle Collaboration using the initial-state radiation method \cite{Belle:2008xmh}, which are also illustrated in Fig. \ref{data} and were plagued by significant uncertainties in both the center-of-mass energy and the cross section. 
In Ref. \cite{Dong:2021juy,Milstein:2022bfg}, the authors suggest that the plateau near the threshold can be understood as the consequence of the Coulomb potential or the Sommerfeld factor, along with the presence of a threshold pole. In Ref. \cite{Dong:2021juy}, the authors stated the pole position can be below threshold. For Ref. \cite{Salnikov:2023qnn}, as a continuation of the work in Ref. \cite{Milstein:2022bfg}, also predicts the existence of a $\Lambda_c\bar{\Lambda}_c$ bound state with a binding energy of 38 MeV. Ref. \cite{Cao:2019wwt} utilizes a model incorporating a Breit-Wigner resonance and $\Lambda_c\bar{\Lambda}_c$ four-point contact interactions to explain the enhancement above the $\Lambda_c\bar{\Lambda}_c$ threshold as a consequence of a virtual pole. Nonetheless, if the $\Lambda_c\bar{\Lambda}_c$
contact coupling were larger, this pole would become a bound state.
This indirectly indicates the experimental evidence for the possible existence of the $\Lambda_c\bar{\Lambda}_c$ bound state.

The existence of the $\Lambda_c\bar{\Lambda}_c$ bound state has been corroborated by numerous theoretical works. Within the one-boson-exchange model \cite{Lee:2011rka,Chen:2017vai}, the $\Lambda_c\bar{\Lambda}_c$ system can exist as bound state and the binding energy of this system is very sensitive to the cutoff. Similarly, the quasipotential Bethe-Salpeter (BS) equation approach also supports the existence of a $\Lambda_c\bar{\Lambda}_c$ bound state \cite{Song:2022svi}. The existence of the $\Lambda_c\bar{\Lambda}_c$ bound state is also supported by the effective field theory \cite{Lu:2017dvm}. Additionally, based on the heavy baryon chiral perturbation theory it is suggested that the $\Lambda_c\bar{\Lambda}_c$ system can form a bound state through two-pion exchange interaction potentials \cite{Chen:2011cta}. In Ref. \cite{Dong:2021juy}, the authors state that the binding energy of this system may range from a few to several tens of MeV depending on the cutoff within the BS equation approach.

 As a formally exact equation to describe the relativistic bound system, the BS equation is formulated in Minkowski space based on the relativistic quantum theory \cite{Salpeter:1951sz,Nakanishi:1969ph}. Over the past decades, this formalism has been successfully used to investigate heavy mesons, heavy baryons, and exotic states \cite{Guo:1996jj,Jain:1993qh,Jin:1992mw,Miransky:2000bd,Zhao:2021cvg,Qi:2021iyv}. In this work we will establish the BS equation for the $\Lambda_c\bar{\Lambda}_c$ system. The interaction kernel will be derived from the four-point Green function with the relevant Lagrangians. Since the strong interaction vertices are determined by the physical particles and the off-shell exchanged particles, a form factor is introduced to account for the finite size effects of the interacting hadrons. Subsequently, the BS equations will be numerically solved under the covariant instantaneous approximation. Moreover, we will explore some possible partial decay widths of the $\Lambda_c\bar{\Lambda}_c$ bound state.

The remainder of this paper is organized as follows. Section \ref{BS} will establish the BS equation and the normalization conditions for the $\Lambda_c\bar{\Lambda}_c$ system. In Sec. \ref{Decay formalism}, the partial decay widths of the $\Lambda_c\bar{\Lambda}_c$ bound states to various final states will be investigated. The numerical results will be presented in Sec. \ref{Num}. The final section will offer our summary.

\section{Bethe-Salpeter equation for the $\Lambda_c\bar{\Lambda}_c$ system}
\label{BS}

As we discussed in Introduction, we will study the possible $\Lambda_c\bar{\Lambda}_c$ bound state. The $S$-wave $\Lambda_c\bar{\Lambda}_c$ system may form two bound states with quantum numbers $J^{PC}=0^{-+}$ and $1^{--}$. The BS wave function for the $\Lambda_c\bar{\Lambda}_c$ bound state is defined as: 
\begin{equation}
\chi_P(x_1,x_2,P)=\langle0|T\Lambda_c(x_1)\bar{\Lambda}_c(x_2)|P\rangle,
\end{equation}
where $P$ ($=Mv$) is the total momentum of the $\Lambda_c\bar{\Lambda}_c$ bound state and $v$ represents its velocity. The BS wave function in the momentum space is defined as
\begin{equation}
\chi_P(x_1,x_2,P)=e^{-iP X}\int\frac{d^4p}{(2\pi)^4}\chi_P(p)e^{-ip x},
\end{equation}
where $X=\lambda_1x_1+\lambda_2x_2$ and $x=x_1-x_2$ are the center-of-mass coordinate and the relative coordinate of the $\Lambda_c\bar{\Lambda}_c$ bound state with $\lambda_{1(2)}=\frac{m_{1(2)}}{m_{1}+m_{2}}=\frac{1}{2}$ and $m_{1(2)}$ is the mass of $\Lambda_c$ ($\bar{\Lambda}_c$), and $p$ is the relative momentum of the bound state. The momenta of $\Lambda_c$ and $\bar{\Lambda}_c$ can be expressed in terms of the relative momentum $p$ and the total momentum $P$ as $p_1=\lambda_1P+p$ and $p_2=\lambda_2P-p$, respectively.

The BS equation in the momentum space can be written as follows \cite{Guo:2007qu,Weng:2010rb}
\begin{equation}\label{BS Eq.}
    \chi_P(p)=S(p_1)\int\frac{d^4q}{(2\pi)^4}\bar{K}(P,p,q)\chi_P(q)S(-p_2),
\end{equation}
where $\bar{K}(P,p,q)$ is the interaction kernel from the irreducible Feynman diagrams, $S(p_1)$ and $S(-p_2)$ are the propagators of $\Lambda_c$ and $\bar{\Lambda}_c$, respectively. For convenience, we define $p_l\equiv v\cdot p$ as the longitudinal projection of $p$ along $v$ and $p_t\equiv p-p_l v$ which is transverse to $v$.

In general, the normalization condition of the BS wave function for the $\Lambda_c$ and $\bar{\Lambda}_c$ system is 
\begin{equation}
    i\int\frac{d^4p}{(2\pi)^4}\chi_P(p)\frac{d^4q}{(2\pi)^4}\frac{\partial}{\partial P^0}\left[I(P,p,q)+\bar{K}(P,p,q)\right]\chi_P(q)=2E_{\mathbf{P}},\,\,P^0=E_{\mathbf{P}},
\end{equation}
where $I(P,p,q)=(2\pi)^4\delta^4(p-q)S^{-1}(p_1)S^{-1}(-p_2)$.

In the heavy quark limit, the propagators of $\Lambda_c$ and $\bar{\Lambda}_c$ can be expressed as the following forms:
\begin{equation}\label{propagator one}
       S(p_1) =i\frac{m_1(1+\slashed{v})}{2w_1(\lambda_1M+p_l-w_1+i\epsilon)},
\end{equation}
and
\begin{equation}\label{propagator two}
       S(p_2)=i\frac{m_2(1+\slashed{v})}{2w_2(\lambda_2M-p_l-w_2+i\epsilon)},
\end{equation}
where the energy $w_{1(2)}=\sqrt{m^2_{1(2)}-p_t^2}$, and $\epsilon$ is the infinitesimal.

Substituting Eqs. (\ref{propagator one}) and (\ref{propagator two}) into Eq. (\ref{BS Eq.}), we obtain the following two constraint relations for the BS wave function $\chi_P(p)$:
\begin{equation}
    \slashed{v}\chi_P(p)=\chi_P(p),
\end{equation}
\begin{equation}
    \chi_P(p)\slashed{v}=-\chi_P(p).
\end{equation}
Then taking into account these two constraint relations and other restrictions from Lorentz covariance and parity transformation on the form of $\chi_P(p)$, the BS wave functions for the $S$-wave pseudoscalar ($J^{PC}=0^{-+}$) and vector ($J^{PC}=1^{--}$) $\Lambda_c\bar{\Lambda}_c$ bound state can be parametrized in the following forms, respectively:
\begin{equation}
    \chi_P(p)=(1+\slashed{v})\gamma_5f_1(p),
\end{equation}
and
\begin{equation}
    \chi^{(r)}_P(p)=(1+\slashed{v})\slashed{\xi}^{(r)}f_2(p),
\end{equation}
where $f_1(p)$ and $f_2(p)$ are the Lorentz-scalar functions of $p_t^2$ and $p_l$, and  $\xi^{(r)}_\mu$ is the polarization vector of the vector $\Lambda_c\bar{\Lambda}_c$ bound state.

To simplify the BS equation (\ref{BS Eq.}), we impose the so-called covariant instantaneous approximation in the kernel: $p_l=q_l=0$. In this approximation the projection of the momentum of each constituent particle along the total momentum $P$ is not changed, the energy exchanged between the constituent particles of the binding system is neglected. This approximation is appropriate since we consider the binding energy of $\Lambda_c\bar{\Lambda}_c$ bound state to be in the range (0-50) MeV, which is very small compared to the constituent particle masses. Under this approximation, the kernel in the BS equation is reduced to $\bar{K}(P,p_t,q_t)$, which will be used in the following calculations.

After some algebra, we find that the BS scalar wave functions for both the pseudoscalar $\Lambda_c\bar{\Lambda}_c$ bound state ($f_1(p)$) and the vector $\Lambda_c\bar{\Lambda}_c$ bound state ($f_2(p)$) satisfy the same integral equation as follows:
\begin{equation}\label{BS scalar WF}
    f(p)=\frac{-m_1m_2}{w_1w_2(\lambda_1M+p_l-w_1+i\epsilon)(\lambda_2M-p_l-w_2+i\epsilon)}\int\frac{d^4q}{(2\pi)^4}\bar{K}(P,p_t,q_t)f(q).
\end{equation}
We integrate both sides of above equation with respect to $p_l$ to obtain
\begin{equation}\label{3D-BS scalar WF}
    \tilde{f}(p_t)=\frac{-im_1m_2}{w_1w_2(M-w_1-w_2)}\int\frac{d^3q_t}{(2\pi)^3}\bar{K}(P,p_t,q_t)\tilde{f}(q_t),
\end{equation}
where we define $\tilde{f}(p_t)=\int dp_l f(p)$.

For later convenience we also write out $f(p)$ in term of $\tilde{f}(p_t)$. From Eqs. (\ref{BS scalar WF}) and (\ref{3D-BS scalar WF}) we have 
\begin{equation}
    f(p)=-i\frac{M-w_1-w_2}{2\pi(\lambda_1M+p_l-w_1+i\epsilon)(\lambda_2M-p_l-w_2+i\epsilon)}\tilde{f}(p_t).
\end{equation}

As the $\Lambda_c$ is an isoscalar state, the total interaction kernel arises from the exchanges of $\omega$ and $\sigma$. The related Lagrangians, constructed with heavy quark and chiral symmetries \cite{Casalbuoni:1996pg,Song:2022yfr}, are presented as follows:
\begin{equation}
    \begin{split}
        \mathcal{L}_{\Lambda_c\Lambda_c\omega}&=-i\frac{g_V\beta_B}{4m_{\Lambda_c}}\omega^\mu\bar{\Lambda}_c\overleftrightarrow{\partial}_\mu\Lambda_c,\\
        \mathcal{L}_{\Lambda_c\Lambda_c\sigma}&=i\ell_B\sigma\bar{\Lambda}_c\Lambda_c,
    \end{split}
\end{equation}
where the adopted coupling constants are $g_V$ = 5.8, $\beta_B$ = 0.87, and $\ell_B$ = -3.1. 

Then the interaction kernel can be derived in the lowest-order form as follows:
\begin{equation}
    \begin{split}
        \bar{K}_\omega(P,p,q)&=(2\pi)^4\delta^4(q_1+q_2-p_1-p_2)\left(\frac{g_V\beta_B}{4m_{\Lambda_c}}\right)^2(p_1+q_1)^\mu(p_2+q_2)^\nu\Delta_{\mu\nu}^\omega(k),\\
        \bar{K}_\sigma(P,p,q)&=(2\pi)^4\delta^4(q_1+q_2-p_1-p_2)\ell_B^2\Delta^\sigma(k),
    \end{split}
\end{equation}
where $\Delta_{\mu\nu}^\omega(k)$ and $\Delta^\sigma(k)$ are the propagators for the exchanged $\omega$ and $\sigma$ mesons, respectively, and $k$ represents the momentum of the exchanged meson. 

To take into account the structure and finite size effect of the interacting hadrons, it is necessary to introduce the form factor at the vertices. For $t$-channel vertices, we use the following monopole and exponential form factors:
\begin{equation}
    F(k^2)=\frac{\Lambda^2-m^2}{\Lambda^2-k^2},
\end{equation}
and 
\begin{equation}
    F(k^2)=e^{(k^2-m^2)/\Lambda^2},
\end{equation}
respectively, where $m$ and $k$ represent the mass and momentum of the exchanged meson. The cutoff parameter $\Lambda$ can be further reparameterized as $\Lambda=m+\alpha \Lambda_{\text{QCD}}$ with $\Lambda_{\text{QCD}}$ = 220 MeV, and the parameter $\alpha$ being of order unity. The value of $\alpha$ depends on the exchanged and external particles involved in the strong interaction vertex and cannot be obtained from the first principle.

\section{The partial decay widths of the $\Lambda_c\bar{\Lambda}_c$ bound state}
\label{Decay formalism}
In this section, we will investigate the decay widths of the $S$-wave $\Lambda_c\bar{\Lambda}_c$ bound states, thereby providing theoretical support for the experimental search for $\Lambda_c\bar{\Lambda}_c$ bound states. For the $\Lambda_c\bar{\Lambda}_c$ bound state with quantum numbers $J^{PC}=1^{--}$, some possible strong decay channels include $N\bar{N}$, $D\bar{D}$, $D\bar{D}^\ast$, $D^\ast\bar{D}^\ast$, $\pi\bar{\pi}$, and $K\bar{K}$. Due to parity constraints, final states such as $D\bar{D}$, $\pi\bar{\pi}$, and $K\bar{K}$ are forbidden for the $\Lambda_c\bar{\Lambda}_c$ bound state with $J^{PC}=0^{-+}$, while only $N\bar{N}$, $D\bar{D}^\ast$, and $D^\ast\bar{D}^\ast$ final states are allowed. 

The strong decays of $S$-wave $\Lambda_c\bar{\Lambda}_c$ bound states into $N\bar{N}$, $D^{(\ast)}\bar{D}^{(\ast)}$, $\pi\bar{\pi}$, and $K\bar{K}$ can occur via exchanging $D/D^\ast$, $N$ (nucleon), $\Sigma_c$, and $\Xi'_c$, respectively. The relevant interaction vertices are \cite{Xie:2015zga,Dong:2014ksa,Huang:2016ygf,Guo:2016iej}
\begin{equation}
    \begin{split}
\mathcal{L}_{DN\Lambda_c}=&ig_{DN\Lambda_c}\bar{\Lambda}_c\gamma^5DN+\rm{H.c.},\\
        \mathcal{L}_{D^\ast N\Lambda_c}=&g_{D^\ast N\Lambda_c}\bar{\Lambda}_c\gamma^\mu ND^\ast_\mu+\rm{H.c.},\\
        \mathcal{L}_{\Lambda_c\Sigma_c\pi}=&g_{\Lambda_c\Sigma_c\pi}\bar{\Lambda}_c\partial^\mu\pi\gamma_5\gamma_\mu\Sigma_c+\rm{H.c.},\\
        \mathcal{L}_{\Lambda_c\Xi'_cK}=&g_{\Lambda_c\Xi'_cK}\bar{\Lambda}_c\partial^\mu K\gamma_5\gamma_\mu\Xi'_c+\rm{H.c.},\\
    \end{split}
\end{equation}
where $\rm{H.c.}$ represent the Hermitian conjugate of the previous terms. The coupling constants $g_{\Lambda_cND}=-13.98$ and $g_{\Lambda_cND^\ast}=-5.2$ are determined from flavour-SU(4) symmetry \cite{Dong:2010xv,Liu:2001ce}, $g_{\Lambda_c\Sigma_c\pi} = \frac{g_2}{\sqrt{2}f_\pi}$ and $g_{\Lambda_c\Xi'_cK} = \frac{g_2}{2f_\pi}$ \cite{Yan:1992gz}, where $f_\pi=93$ MeV is the pion decay constant, and $g_2=0.565$ is determined from the $\Sigma^{++}_c\rightarrow\Lambda^+_c\pi^+$ decay \cite{Cheng:2015naa}.

In the rest frame, the two-body decay width of the bound state is expressed as
\begin{equation}
    d\Gamma=\frac{1}{32\pi^2}|\mathcal{M}|^2\frac{|\mathbf{p}'|}{M^2}d\Omega,
\end{equation}
where $\mathcal{M}$ is the Lorentz invariant decay amplitude of the process. $|\mathbf{p}'|$ is the magnitude of the three-momentum of the final state particles in the rest frame of the bound state, defined by
\begin{equation}
    |\mathbf{p}'|=\frac{1}{2M}\sqrt{\lambda(M^2,m_1^2,m_2^2)},
\end{equation}
with $\lambda(a,b,c)=a^2+b^2+c^2-2ab-2ac-2bc$ being the K$\mathrm{\ddot{a}}$ll$\mathrm{\acute{e}}$n function.

The lowest order Lorentz-invariant decay amplitude for the $\Lambda_c\bar{\Lambda}_c$ bound state decaying into $N\bar{N}$ is 
\begin{equation}
    \begin{split} \mathcal{M}_{N\bar{N}}=&\mathcal{M}_{N\bar{N}}^D+\mathcal{M}_{N\bar{N}}^{D^\ast}\\
       =&\int\frac{d^4p}{(2\pi)^4}g_{D N\Lambda_c}^2\bar{u}(p'_1)\gamma^5\chi_P(p)\gamma^5v(p'_2)\Delta_D(k,m_D)F^2(k^2,m_D)\\
       &-\int\frac{d^4p}{(2\pi)^4}g_{D^\ast N\Lambda_c}^2\bar{u}(p'_1)\gamma_\mu\chi_P(p)\gamma_\nu v(p'_2)\Delta_{D^\ast}^{\mu\nu}(k,m_{D^\ast})F^2(k^2,m_{D^\ast}),
    \end{split}
\end{equation}
where $p'_{1(2)}=\left(E_{1(2)},(-)\mathbf{p}'\right)$ denotes the momentum of $N(\bar{N})$ in the final state, $k$ is the momentum transfer in the decay process, $\bar{u}(p'_1)$ and $v(p'_2)$ are the Dirac spinors of $N$ and $\bar{N}$, respectively, $\Delta_D(k,m_D)$ and $\Delta_{D^\ast}^{\mu\nu}(k,m_{D^\ast})$ are the propagators for the exchanged mesons. For convenience, we define $p'\equiv \lambda_2 p'_1-\lambda_1p'_2$ which is not the relative momentum of the final particles, and is given by $p'=(\lambda_2E_1-\lambda_1E_2,\mathbf{p}')$. The BS wave function $\chi_P(p)$ could be either pseudoscalar or vector.

The lowest-order Lorentz-invariant decay amplitudes are 
\begin{equation}
   \mathcal{M}_{D\bar{D}}=\int\frac{d^4p}{(2\pi)^4}g_{DN\Lambda_c}^2\gamma_5\Delta_{N}(k,m_N)\gamma_5\chi_P(p)F^2(k^2,m_{N}),
\end{equation}
\begin{equation}
   \mathcal{M}_{D\bar{D}^\ast}=-i\int\frac{d^4p}{(2\pi)^4}g_{D N\Lambda_c}g_{D^\ast N\Lambda_c}\epsilon^\ast_\mu(p'_2)\gamma_5\chi_P(p)\gamma^\mu\Delta_{N}(k,m_N)F^2(k^2,m_{N}),
\end{equation}
and 
\begin{equation}
   \mathcal{M}_{D^\ast\bar{D}^\ast}=-\int\frac{d^4p}{(2\pi)^4}g_{D^\ast N\Lambda_c}^2\epsilon^\ast_\mu(p'_1)\gamma^\mu\chi_P(p)\gamma^\nu\epsilon_\nu^\ast(p'_2)\Delta_{N}(k,m_N)F^2(k^2,m_{N}),
\end{equation}
for the $D\bar{D}$ (only allowed for the vector $\Lambda_c\bar{\Lambda}_c$ bound state), $DD^\ast$, and $D^\ast\bar{D^\ast}$ final states, respectively, $\epsilon$ is the polarization vector of $D^\ast$ or $\bar{D}^\ast$.

For the $\pi\bar{\pi}$ and $K\bar{K}$ final states, the lowest-order Lorentz-invariant decay amplitudes are given by
\begin{equation}
   \mathcal{M}_{\pi\bar{\pi}}=\int\frac{d^4p}{(2\pi)^4}g_{\Lambda_c\Sigma_c\pi}^2\gamma_\mu\gamma^5\chi_P(p)\gamma_5\gamma_\nu p^{'\mu}_1p^{'\nu}_2\Delta_{\Sigma_c}(k,m_{\Sigma_c})F^2(k^2,m_{\Sigma_c}),
\end{equation}
and 
\begin{equation}
   \mathcal{M}_{K\bar{K}}=\int\frac{d^4p}{(2\pi)^4}g_{\Lambda_c\Xi'_cK}^2\gamma_\mu\gamma^5\chi_P(p)\gamma_5\gamma_\nu p^{'\mu}_1p^{'\nu}_2\Delta_{\Xi'_c}(k,m_{\Xi'_c})F^2(k^2,m_{\Xi'_c}),
\end{equation}
respectively. These two decay processes are exclusive to the vector $\Lambda_c\bar{\Lambda}_c$ bound state.

\section{Numerical results}
\label{Num}
To show the numerical results, we begin by presenting the masses of mesons and baryons in Table \ref{Mass} \cite{ParticleDataGroup:2022pth}, which are essential for studying the $\Lambda_c\bar{\Lambda}_c$ bound states and their potential decay channels. Our model includes a single free parameter, $\alpha$, which is influenced by the exchanged particle and the external particles at the strong interaction vertex. It is expected that $\alpha$ is order unity and cannot be decided from the first principle. Consequently, the binding energy $E_b$ (defined as $E_b=m_1+m_2-M$, where we consider $\Lambda_c\bar{\Lambda}_c$ as a shallow bound state system with $E_b$ ranging from 0 to 50 MeV) of the $\Lambda_c\bar{\Lambda}_c$ system cannot be determined, as it depends on the value of the parameter $\alpha$. To explore possible solutions for the $\Lambda_c\bar{\Lambda}_c$ bound states, we vary $\alpha$ over a broader range (0.5-5). 

\begin{table}[h]
\renewcommand{\arraystretch}{1.2}
\centering
\caption{Masses (unit: MeV) of the relevant mesons and baryons.}\label{Mass}
\begin{tabular*}{\textwidth}{@{\extracolsep{\fill}}cccccccccc}
\hline
\hline
$\Lambda_c^+$ & $\Sigma_c^{++}$  & $\Sigma_c^{+}$ & $\Sigma_c^{0}$ & $\Xi^{'+}_c$ & $\Xi^{'0}_c$ & $p$ & $n$ \\ 
\hline
2286.46  & 2453.97 & 2452.65 & 2453.75 & 2578.2 & 2578.70 & 938.27 & 939.57 \\
\hline\hline
 $D^0$ & $D^\pm$ & $D^{\ast0}$ & $D^{\ast\pm}$ & $\pi^0$ & $\pi^\pm$ & $K^0$ & $K^\pm$ & $\omega$ & $\sigma$\\
 \hline
 1864.84 & 1869.66 & 2006.85 & 2010.26 & 134.98 & 139.57 & 497.61 & 493.68 & 782.66 & 500\\
\hline
\hline
\end{tabular*}\label{coupling constants}
\end{table}

\begin{figure}[ht]
\centering
    \rotatebox{0}{\includegraphics*[width=0.5\textwidth]{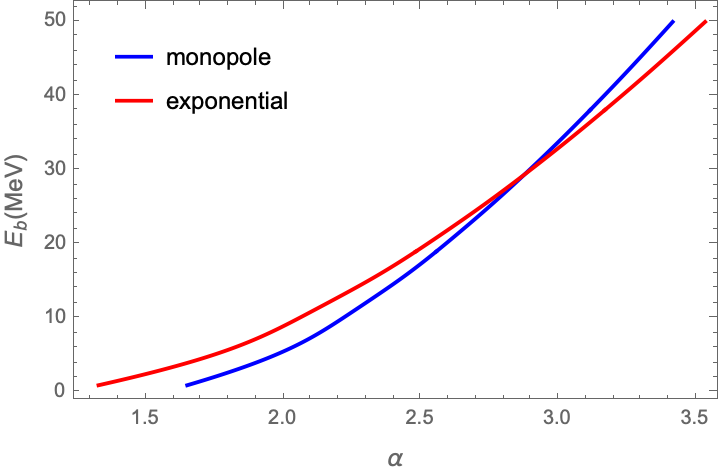}}
    \caption{Values of $\alpha$ and $E_b$ for the possible bound states for the $\Lambda_c\bar{\Lambda}_c$ system.}
  \label{boundstate}
\end{figure}

\begin{figure}[htbp]
\centering
\subfigure[]{
\includegraphics[width=8cm]{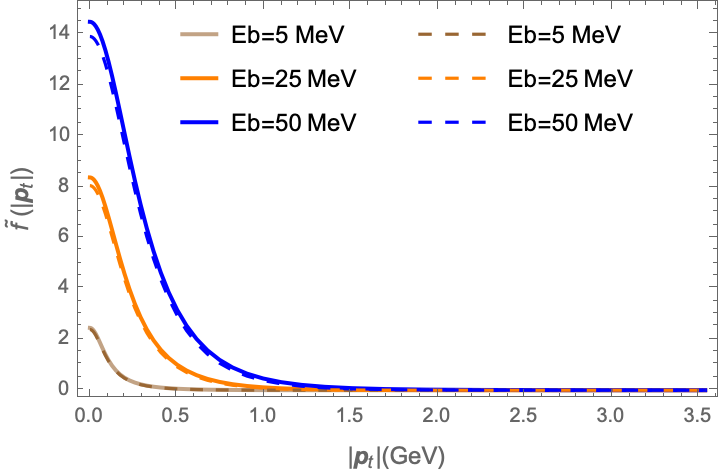}
}
\,
\subfigure[]{
\includegraphics[width=8cm]{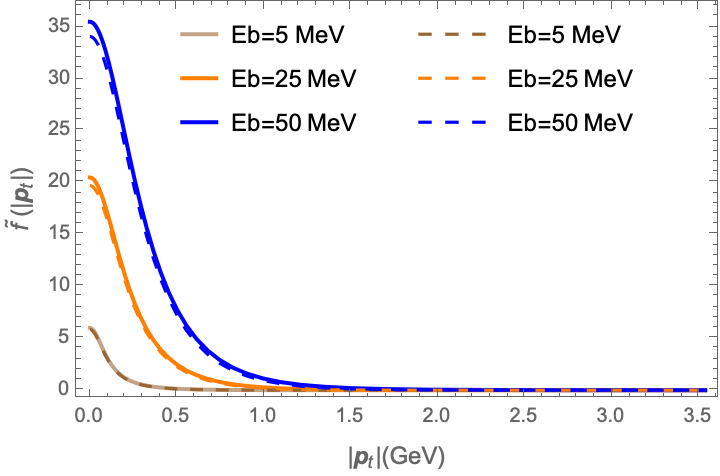}
}
\caption{Numerical results of the normalized scalar equation $\tilde{f}(|\mathbf{p}_t|)$ for (a) pseudoscalar and (b) vector $\Lambda_c\bar{\Lambda}_c$ bound states. The solid and dashed curves correspond to the monopeole form factor and the exponential form factor, respectively.}
\label{Num wave function}
\end{figure}

To solve the integral equation (\ref{3D-BS scalar WF}), we discretize the integration region into $n$ pieces (with $n$ being sufficiently large), transforming Eq. (\ref{3D-BS scalar WF}) to an eigenvalue equation for the $n$ dimensional vector $\tilde{f}$. After solving the eigenvalue equation, we find that when the parameter $\alpha$ is in the range of 1.65 to 3.41 for the monopole form factor and 1.32 to 3.53 for the exponential form factor, the $\Lambda_c\bar{\Lambda}_c$ system can exist as a bound state with a binding energy in the range of 0-50 MeV. This indicates that the contributions from the exchanges of $\omega$ and $\sigma$ are sufficient to allow the $\Lambda_c\bar{\Lambda}_c$ system to form bound states. The values of $\alpha$ and the corresponding binding energy $E_b$ are depicted in Fig. \ref{boundstate}, indicating that the binding energy $E_b$ increases with parameter $\alpha$. This trend is attributed to the fact that as the parameter $\alpha$ increases, the effective range and intensity of the interactions between the constituent particles of the bound state are enhanced, resulting in a stronger binding. As we discussed in Sec.\ref{BS}, for both the pseudoscalar and vector $\Lambda_c\bar{\Lambda}_c$ systems, they satisfy the same scalar BS equation (\ref{BS scalar WF}), and therefore, they exhibit the same trend but with different normalization factors. In Fig. \ref{Num wave function}, we present the numerical results of the normalized scalar BS wave functions for the pseudoscalar and vector $\Lambda_c\bar{\Lambda}_c$ bound states with binding energies $E_b$ = 5 MeV, 25 MeV, and 50 MeV.

\begin{figure}[htbp]
\centering
\subfigure[]{
\includegraphics[width=5.4cm]{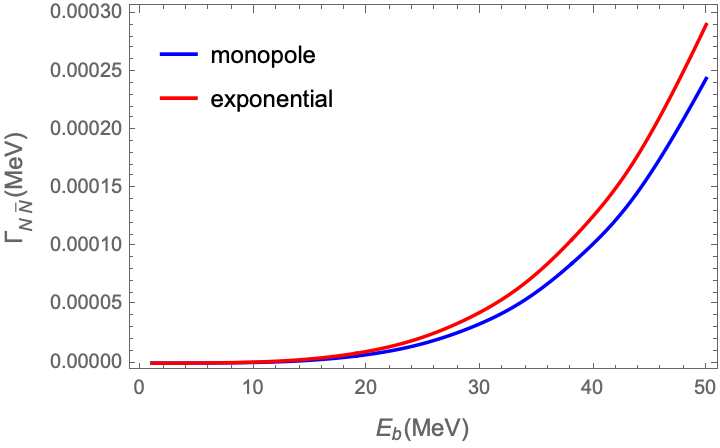}
}
\,
\subfigure[]{
\includegraphics[width=5.4cm]{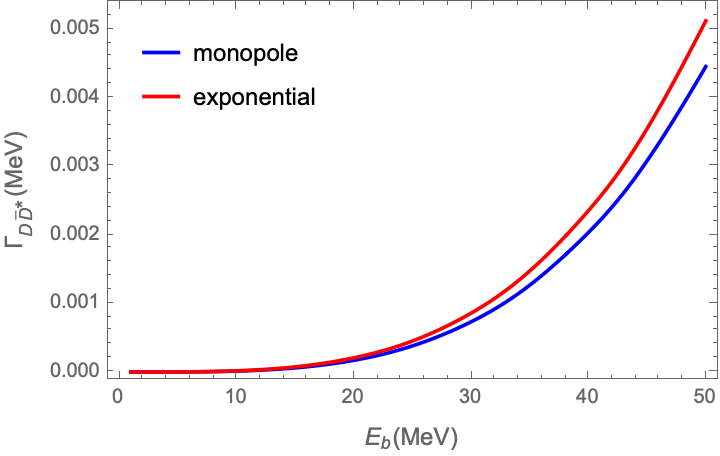}
}
\,
\subfigure[]{
\includegraphics[width=5.4cm]{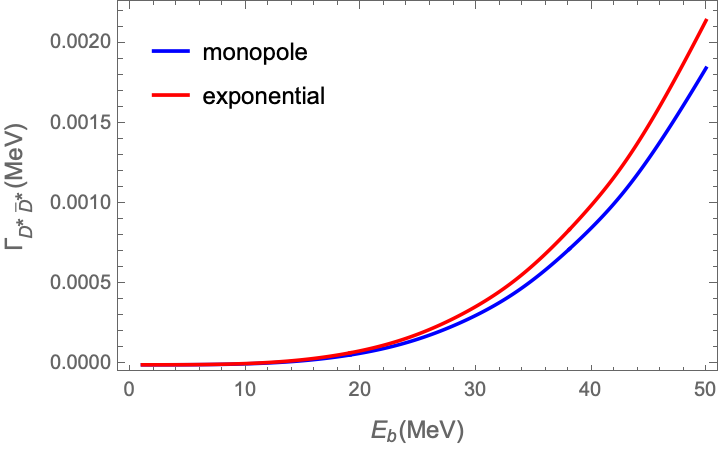}
}
\caption{The particle decay widths of the $J^{PC}=0^{-+}$ $\Lambda_c\bar{\Lambda}_c$ bound state to (a) $N\bar{N}$, (b) $D\bar{D}^\ast$, and (c) $D^\ast\bar{D}^\ast$ final states, respectively.}
\label{SWidth}
\end{figure}

Taking into account the constraints of quantum numbers such as spin and parity, the $\Lambda_c\bar{\Lambda}_c$ bound states with $J^{PC}=0^{-+}$ can decay through strong interactions into $N\bar{N}$, $D\bar{D}^\ast$, and $D^\ast\bar{D}^\ast$ final states. The estimated partial decay widths for these final states are $1.40\times10^{-10} \sim 2.45\times10^{-4}$ MeV, $4.01\times10^{-9} \sim 4.46\times10^{-3}$ MeV, and $4.01\times10^{-9} \sim 4.46\times10^{-3}$ MeV for monopole form factors, $2.88\times10^{-10} \sim 2.91\times10^{-4}$ MeV, $6.07\times10^{-9} \sim 5.13\times10^{-3}$ MeV, and $2.60\times10^{-9} \sim 2.15\times10^{-3}$ MeV for exponential form factors, with the binding energy ranging from 0 to 50 MeV. The partial decay widths are also illustrated in Fig. \ref{SWidth}.  The decay to the $N\bar{N}$ final state is suppressed by the OZI rule due to $c\bar{c}$ annihilation, resulting in the smallest decay width among the considered channels. The decays of the $\Lambda_c\bar{\Lambda}_c$ bound state with $J^{PC}=0^{-+}$ to both $D\bar{D}^\ast$ and $D^\ast\bar{D}^\ast$ final states are mediated by $N$ exchange. Due to the stronger coupling at the $\Lambda_cDN$ vertex and the larger phase space for the $D\bar{D}^\ast$ final state, the decay width of the $D\bar{D}^\ast$ final state is larger than that of the $D^\ast\bar{D}^\ast$ final state.

The partial decay widths of the $J^{PC}=1^{--}$ $\Lambda_c\bar{\Lambda}_c$ bound state to $N\bar{N}$, $D\bar{D}$, $D\bar{D}^\ast$, $D^\ast\bar{D}^\ast$, $\pi\bar{\pi}$, and $K\bar{K}$ final states range from $1.86\times10^{-10}\sim8.31\times10^{-4}$ MeV, $1.12\times10^{-7}\sim0.18$ MeV, $4.00\times10^{-4}\sim16.85$ MeV, $1.85\times10^{-8}\sim1.68\times10^{-2}$ MeV, $1.41\times10^{-14}\sim6.64\times10^{-8}$ MeV, and $4.49\times10^{-13}\sim2.23\times10^{-6}$ MeV for monopole form factors, and $7.02\times10^{-10}\sim1.60\times10^{-3}$ MeV, $1.10\times10^{-7}\sim0.19$ MeV, $7.04\times10^{-4}\sim19.25$ MeV, $1.89\times10^{-8}\sim2.03\times10^{-2}$ MeV, $8.77\times10^{-14}\sim1.73\times10^{-7}$ MeV, and $3.16\times10^{-12}\sim6.26\times10^{-6}$ MeV for exponential form factors, with the binding energy ranging from 0 to 50 MeV. The $\Lambda_c\bar{\Lambda}_c$ bound state with $J^{PC}=1^{--}$ exhibits larger decay widths compared to the state with $J^{PC}=0^{-+}$. The decay processes to $\pi\bar{\pi}$ and $K\bar{K}$ final states, involving the $c\bar{c}$ annihilation, are suppressed by the OZI rule, and the coupling strengths at the $\Lambda_c\Sigma_c\pi$ and $\Lambda_c\Xi'_cK$ vertices are relatively small, leading to minimal decay widths as anticipated. The $D\bar{D}^\ast$ final state still has the largest decay width in the decay of the $J^{PC}=1^{--}$ $\Lambda_c\bar{\Lambda}_c$ bound state.

\begin{figure}[htbp]
\centering
\subfigure[]{
\includegraphics[width=5.5cm]{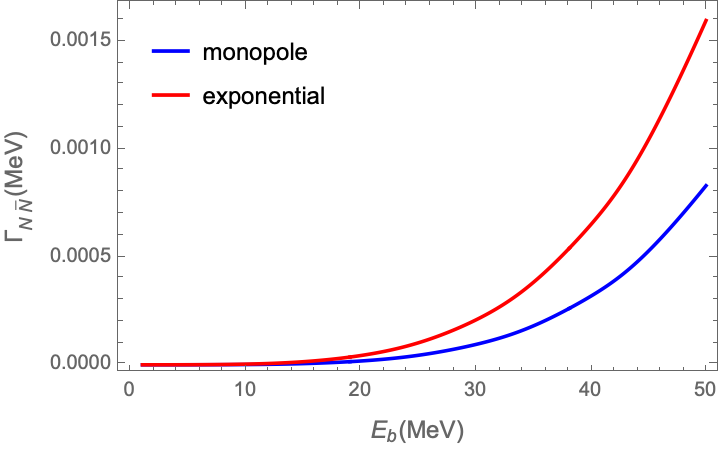}
}
\,
\subfigure[]{
\includegraphics[width=5.3cm]{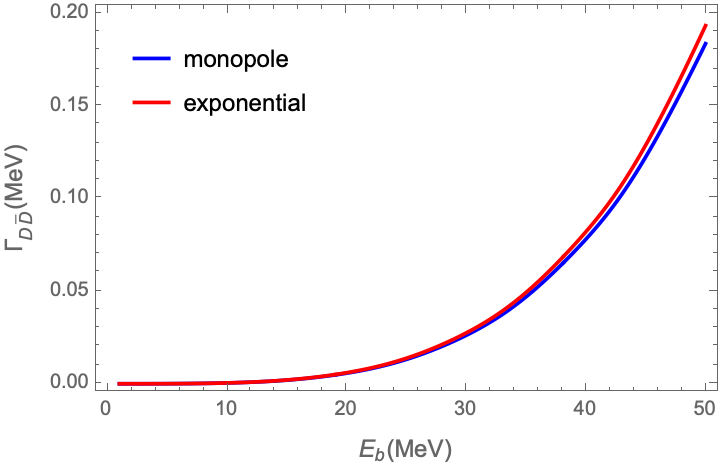}
}
\,
\subfigure[]{
\includegraphics[width=5.2cm]{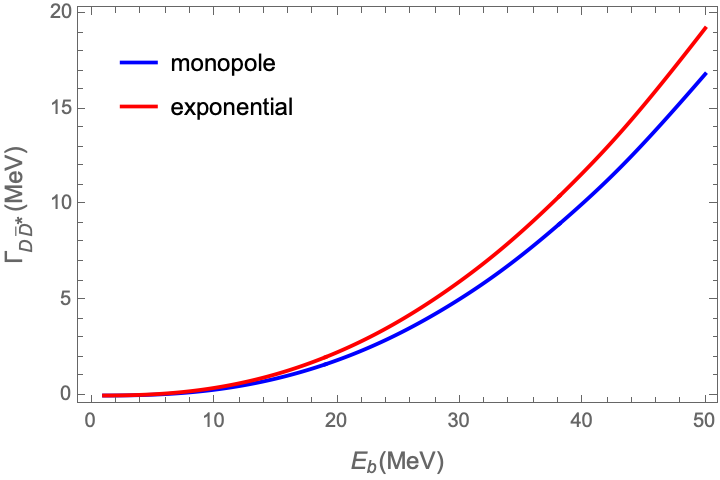}
}
\,
\subfigure[]{
\includegraphics[width=5.2cm]{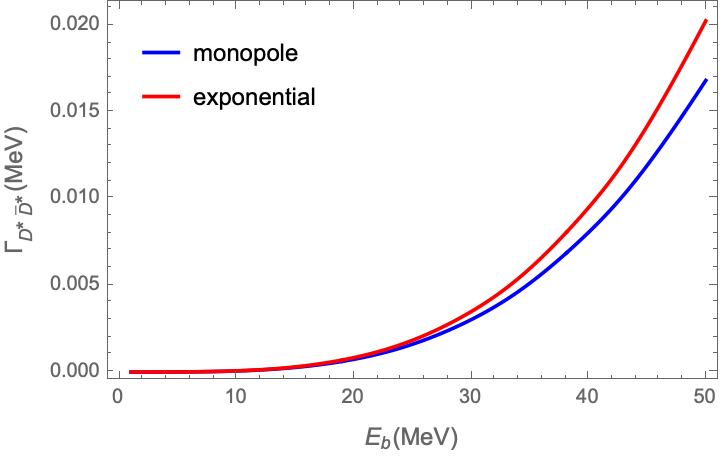}
}
\,
\subfigure[]{
\includegraphics[width=5.5cm]{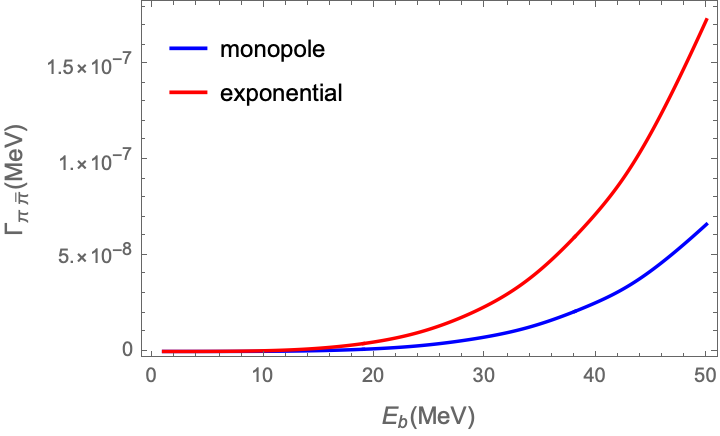}
}
\,
\subfigure[]{
\includegraphics[width=5.4cm]{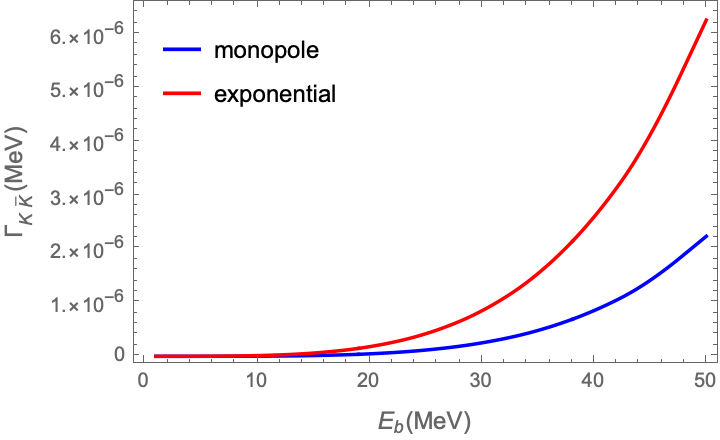}
}
\caption{The particle decay widths of the $J^{PC}=1^{--}$ $\Lambda_c\bar{\Lambda}_c$ bound state to (a) $N\bar{N}$, (b) $D\bar{D}$, (c) $D\bar{D}^\ast$, (d) $D^\ast\bar{D}^\ast$, (e) $\pi\bar{\pi}$, and (f) $K\bar{K}$ final states, respectively.}
\label{VWidth}
\end{figure}

As presented in Figs. \ref{SWidth} and \ref{VWidth}, the partial decay widths of the $\Lambda_c\bar{\Lambda}_c$ bound states exhibit a similar trend of increasing with the binding energy $E_b$, which is determined by the parameter $\alpha$ as shown in Fig. \ref{boundstate}. Intuitively, one might expect smaller decay widths with increased binding due to reduced phase space. However, as $\alpha$ increases, the corresponding cutoff parameter $\Lambda$ also increases. This causes the value of the form factor to increase at low momentum transfer and only decrease significantly at high momentum transfer, leading to larger decay widths. Additionally, the normalized scalar BS wave function $\tilde{f}(|\mathbf{p}_t|)$ also increases significantly with binding energy, as shown in Fig. \ref{Num wave function}. Therefore, the decay widths increase with binding energy rather than decrease.

If we adopt the binding energy $E_b = 38$ MeV for the $J^{PC} = 1^{--}$ $\Lambda_c\bar{\Lambda}_c$ bound state as predicted in Ref. \cite{Salnikov:2023qnn}, the partial decay widths for the $N\bar{N}$, $D\bar{D}$, $D\bar{D}^\ast$, $D^\ast\bar{D}^\ast$, $\pi\bar{\pi}$, and $K\bar{K}$ final states are $2.56\times10^{-4}$ MeV, $6.43\times10^{-2}$ MeV, 8.92 MeV, $6.78\times10^{-3}$ MeV, $2.05\times10^{-8}$ MeV, and $6.83\times10^{-7}$ MeV for monopole form factors with $\alpha$ = 3.11, respectively, and $5.47\times10^{-4}$ MeV, $6.97\times10^{-2}$ MeV, 10.37 MeV, $8.19\times10^{-3}$ MeV, $6.06\times10^{-8}$ MeV, and $2.19\times10^{-6}$ MeV for exponential form factors with $\alpha$ = 3.16, respectively. These results show that the $D\bar{D}^\ast$ final state predominates in the decay of the $J^{PC} = 1^{--}$ $\Lambda_c\bar{\Lambda}_c$ bound state. Therefore, we propose searching for the $\Lambda_c\bar{\Lambda}_c$ bound state in the $D\bar{D}^\ast$ final state.

In addition to the two-body strong decay processes we have studied, three-body decays are also important for the $\Lambda_c\bar{\Lambda}_c$ bound state because these processes do not involve quark-antiquark annihilation. For example, decays to $\eta_c\pi\pi$ and $J/\psi\pi\pi$ are significant. In Ref. \cite{Qiao:2005av}, the authors argue that $Y(4260)$ is a deeply bound $\Lambda_c\bar{\Lambda}_c$ state and propose that $J/\psi\pi\pi$ is its dominant mode of decay, and that there are enough events to observe $Y(4260)$ in the $\psi'\pi\pi$ channel. In Ref. \cite{Wan:2021vny}, the authors studied baryon-antibaryon molecular states and similarly suggested that important decay modes for $\Lambda\bar{\Lambda}$ molecular states include $\eta\pi\pi$, $\omega\pi\pi$, and others, since these processes involve only quark rearrangements and not quark-antiquark annihilation. Currently, the study of three-body decays faces certain computational challenges in our model. In addition, the coupling parameters of the $\Lambda_c\bar{\Lambda}_c$ bound states to the $\eta_c\pi\pi$ and $J/\psi\pi\pi$ decays are not well understood. We aim to address these challenges in future work to obtain a more comprehensive understanding of the decay processes.

\section{Summary}
\label{Conclusion}

In this study, we have presented a comprehensive analysis of the possible $\Lambda_c\bar{\Lambda}_c$ bound states and some possible strong decays of these bound states. Our theoretical framework is based on the BS equation, which provides a relativistically consistent description of the bound state system. The interaction kernel of the BS equation was constructed from relevant Lagrangians that describe the strong interaction vertices among the exchanged ($\omega$ and $\sigma$) and external ($\Lambda_c$) particles.

Our numerical results indicate that $\Lambda_c\bar{\Lambda}_c$ bound states could exist. However, we cannot determine the mass of the bound state precisely, as it depends on the value of the parameter $\alpha$, which is not determined from first principles and reflects the non-perturbative nature of QCD at low energies. Through our study, we found that when the parameter $\alpha$ is in the range of 1.65 to 3.41 for the monopole form factor and 1.32 to 3.53 for the exponential form factor, the $\Lambda_c\bar{\Lambda}_c$ system can exist as a bound state with the binding energy in the range of 0-50 MeV.

We have also calculated the partial decay widths of the $\Lambda_c\bar{\Lambda}_c$ bound states for various decay channels. For the $J^{PC}=0^{-+}$ state, the decays to $N\bar{N}$, $D\bar{D}^\ast$, and $D^\ast\bar{D}^\ast$ final states were considered. The decay widths increase with the binding energy, with the decay to the $D\bar{D}^\ast$ final state being the most significant one. The OZI rule suppresses the decay to the $N\bar{N}$ final state, resulting in the smallest decay width for this channel.

For the $J^{PC}=1^{--}$ state, we investigated the decay channels to $N\bar{N}$, $D\bar{D}$, $D\bar{D}^\ast$, $D^\ast\bar{D}^\ast$, $\pi\bar{\pi}$, and $K\bar{K}$ final states. The decay widths for the $\pi\bar{\pi}$ and $K\bar{K}$ channels are the smallest due to the suppression by the OZI rule and the relatively small coupling constants at the $\Lambda_c\Sigma_c\pi$ and $\Lambda_c\Xi'_cK$ vertices. Similar to the $J^{PC}=0^{-+}$ state, the decay to the $D\bar{D}^\ast$ final state is the most prominent one for the $J^{PC}=1^{--}$ state.

The results presented in this work have important implications for the experimental search for $\Lambda_c\bar{\Lambda}_c$ bound states. The predicted decay widths can serve as a guide for future experiments at facilities such as BES$\text{\uppercase\expandafter{\romannumeral3}}$, LHCb, and Belle $\text{\uppercase\expandafter{\romannumeral2}}$. Further theoretical and experimental investigations are needed to validate the existence of these bound states and to refine the understanding of their properties.

\acknowledgments
This work was supported by National Natural Science Foundation of China (Project Nos. 12105149 and 12275024).

\end{document}